\documentclass[%
  iicol,
  sn-aps
]{sn-jnl}
\usepackage{xcolor,manyfoot}
\usepackage{amsmath,amssymb}
\usepackage{amscd}
\usepackage{braket}
\usepackage{orcidlink}
\begin{document}
\title[DAGs for Scattering Amplitudes]{%
  Functional Directed~Acyclical~Graphs for
  Scattering~Amplitudes in Perturbation~Theory}
\author{\fnm{Thorsten} \sur{Ohl} \orcidlink{0000-0002-7526-2975}}
\email{ohl@physik.uni-wuerzburg.de}
\affil{%
  \orgname{University of~W\"urzburg},
  \orgdiv{Institute of Theoretical Physics and Astrophysics},
  \orgaddress{%
    \street{Emil-Hilb-Weg 22},
    \postcode{97074}
    \city{W\"urzburg},
    \country{Germany}}}
\abstract{%
  I describe a mathematical framework for the efficient processing of
  the very large sets of Feynman diagrams contributing to the
  scattering of many particles.  I reexpress the
  established numerical methods for the recursive construction of
  scattering elements as operations on compact abstract data
  types.  This allows efficient perturbative
  computations in arbitrary models, as long as they can be described
  by an effective, not necessarily local, Lagrangian.}
\maketitle

\section{Introduction}
The efficient and reliable computation of scattering amplitudes for many
particles in a large class of models,
both on tree level and including higher order corrections,
is a central element of all efforts for analyzing the physics at LHC
and possible future colliders.

Since the first release of \textsc{Madgraph}~\cite{Stelzer:1994ta} about 30
years ago, there has been tremendous progress in the capabilities of the
tools that can compute such scattering amplitudes numerically.  Replacing sums of
Feynman diagrams by recursive numerical evaluation opened the
realm of many-legged amplitudes, including loop corrections.  In
fact, the treatment of QCD corrections has matured so much that tools
like \textsc{Madgraph5}~\cite{Alwall:2011uj} are now employed regularly by
endusers for LHC physics.  Electroweak radiative corrections are
starting to become available in user friendly tools and recursive
techniques are being applied to loop calculations.  At the same time,
de facto standard formats like
UFO~\cite{Degrande:2011ua,Darme:2023jdn} allow the specification of
almost any physics model that might be of interest in the near and not
so near future.

In this paper, I will elaborate a common mathematical structure behind
the recursive calculations.  The focus is not on the immediate
numerical evaluation, but on the elucidation of an algebraic
structure that will later be translated into numerical code.  This
simplifies supporting more general interactions, because purely numerical
codes have to make assumptions that can turn out to be hard to relax
later. In addition, algebraic expressions can be used to generate more
comprehensive tests of models and implementations.  They also simplify
the automatic generation of the additional expressions needed for
subtractions schemes~\cite{Catani:1996vz}.

Finally, at a time when functional programming and strong type systems
are moving more and more from academia into the mainstream, it is a
useful exercise to reconstruct the mathematical structures in a way
that can easily be translated into efficient programs making use of
these paradigms.
The mathematical structures presented here have not been developed
in a vacuum, but are a distillation of commonalities observed in the
concrete data structures implemented for the matrix element generator
\textsc{O'Mega}~\cite{Moretti:2001zz} that is part of the
\textsc{Whizard} event
generator~\cite{Kilian:2007gr}.

Nothing in the following discussion will be specific to
leading order, tree level
matrix elements.  Exactly the same structures appear when implementing loops
using additional
legs~\cite{Cascioli:2011va,Buccioni:2019sur,Actis:2012qn,Actis:2016mpe}
or when adding higher order contributions as terms in an effective
action using a skeleton expansion. The translation of the algebraic
expressions into robust numeric code calling sophisticated external
libraries for loop integrals~\cite{Denner:2016kdg} is much more
challenging, of course. However, also here the algebraic step offers
more options than a purely numerical approach.

The outline of the paper is as follows:
in section~\ref{sec:amplitudes}, I briefly review the recursive
techniques used for computing scattering amplitudes for processes with many
external particles.  This section also serves the purpose of establishing
the terminology and notation used in the remaining sections.  In
section~\ref{sec:DAG}, I introduce Directed Acyclical Graphs~(DAGs),
bundles and their relationships.
In section~\ref{sec:feynman}, I present an algorithm
for efficiently constructing the DAGs representing scattering
amplitudes. In section~\ref{sec:code}, I briefly describe how to
generate efficient numerical code from DAGs constructed according to
the algorithm presented in the previous sections.
In appendix~\ref{app:ocaml}, I sketch the implementation of DAGs and
bundles in \textsc{O'Mega}~\cite{Moretti:2001zz,Kilian:2007gr}.

\section{Scattering Amplitudes}
\label{sec:amplitudes}

It has long been recognized that the textbook representation of
scattering amplitudes as a sum of Feynman diagrams becomes very inefficient
as the number of external particles rises.  Indeed, even though
general estimates are hard to derive for realistic models with
conserved quantum numbers, analytic formulae for toy models and
explicit calculations for specific processes confirm the expectation
that the number of tree level Feynman diagrams grows factorially with
the number of external particles.  If Feynman diagrams with loops are
represented by tree
diagrams~\cite{Cascioli:2011va,Buccioni:2019sur,Actis:2012qn,Actis:2016mpe},
each loop adds two more external particles.
In addition to requiring prohibitive computational resources, the
destructive interferences inherent in gauge theories lead to a loss of
precision if too many terms are added. Starting with $2\to6$ processes
at tree level, the need for a more efficient representation became
evident.

In order to simplify the notation in this section,
I will cross all scattering amplitudes from~$n_{\text{in}}\to
n_{\text{out}}$ to~$n=n_{\text{in}}+n_{\text{out}}\to0$.  Except for
the momentum, I will also suppress all quantum numbers in this
introductory section.  The treatment of general quantum numbers (spin,
flavor, color, etc.) will be the focus of the following sections.

\subsection{Recursion}
\label{sec:recursion}

The appropriate building blocks to replace Feynman diagrams turned out
to be $k$-particle matrix elements of fields
\begin{subequations}
\label{eq:off-shell-nodes}
\begin{equation}
\label{eq:off-shell-phi}
  \phi(\{i_1,i_2,\ldots,i_k\}) =
    \braket{0|\Phi|p_{i_1},p_{i_2},\ldots,p_{i_k}}
\end{equation}
which will be referred to as \emph{wavefunctions} or
of their associated \emph{currents}
\begin{equation}
\label{eq:off-shell-j}
  j(\{i_1,i_2,\ldots,i_k\}) =
    \braket{0|J|p_{i_1},p_{i_2},\ldots,p_{i_k}}\,,
\end{equation}
\end{subequations}
as pioneered by~Berends and Giele~\cite{Berends:1987me}.
The set of indices~$I=\{i_1,i_2,\ldots,i_k\}$ is a subset of the indices
enumerating the external particles or open loop momenta.

Since $I\in 2^{\{1,2,\ldots,n\}}$, the number of possible different
wavefunctions or currents grows only as an exponential~$2^n$ instead of a
factorial~$n!\sim n^n$.  Furthermore, both can be computed recursively
\begin{subequations}
\label{eq:recursion}
  \begin{align}
    \phi(I) &= \sum_{I_1\cup I_2=I} P_I V_{I,I_1,I_2} \phi(I_1)\phi(I_2) \\
       j(I) &= \sum_{I_1\cup I_2=I} V_{I,I_1,I_2} P_{I_1}j(I_1)P_{I_2}j(I_2)\,,
  \end{align}
\end{subequations}
\emph{without} expanding them into Feynman diagrams, which would
reintroduce factorial growth.  In~\eqref{eq:recursion}, $P_I$ denotes
a propagator and~$V_{I,I_1,I_2}$ a vertex factor for three legs.  The
generalization to models containing vertices with more than three legs is obvious.

Note that~$\phi$ is just~$j$ multiplied by a momentum space
propagator.  Thus the choice between the two is only a matter of
convenience.  The rest of the paper will mostly refer to
wavefunctions~\eqref{eq:off-shell-phi}, but
all constructions can be repeated trivially for the currents~\eqref{eq:off-shell-j}.

\subsection{Topologies}
\label{sec:topologies}

There are many ways in which a scattering amplitude~$\mathcal{M}$ can
be constructed from~\eqref{eq:off-shell-nodes}.  The first approach
observes that the~$j$
in~\eqref{eq:off-shell-j} is already amputated. It therefore suffices to set
the momentum
\begin{equation}
  p_1 = - \sum_{i=2}^n p_i
\end{equation}
on the mass shell of particle~$1$ to obtain the scattering amplitude
using the LSZ prescription
\begin{subequations}
\label{eq:M}
\begin{equation}
\label{eq:Helac}
  \mathcal{M}(\{1,2,\ldots,n\}) = j(\{2,3,\ldots,n\})\,.
\end{equation}
This is implemented numerically in
\textsc{Helac}~\cite{Kanaki:2000ey,Cafarella:2007pc} and
\textsc{Recola}~\cite{Actis:2012qn,Actis:2016mpe}.

The second approach
glues the~$\phi$ from~\eqref{eq:off-shell-phi} at vertices to
obtain the scattering amplitude in the form
\begin{multline}
\label{eq:Keystones}
  \mathcal{M}(\{1,2,\ldots,n\}) =\\
    \sum_{I_1\cup I_2\cup I_3=\{1,\ldots,n\}} K_{I_1,I_2,I_3} \phi(I_1)\phi(I_2)\phi(I_3)
\end{multline}
with obvious generalizations to models containing vertices with more
than three legs.  The partitions~$(I_1, I_2, I_3)$ of the external
particles must be chosen carefully to avoid double
counting~\cite{Caravaglios:1995cd,Moretti:2001zz}
and the \emph{keystone}s~$K$ correspond to vertex factors.
This approach was pioneered for numerical calculations in the
standard model by
\textsc{Alpha}~\cite{Caravaglios:1995cd,Caravaglios:1996nq,Caravaglios:1998yr}
and is implemented as an algebraic algorithm for arbitrary models in
\textsc{O'Mega}~\cite{Moretti:2001zz,Kilian:2007gr}.

The third approach combines
the DAGs at propagators
\begin{equation}
\label{eq:Comix}
  \mathcal{M}(\{1,2,\ldots,n\}) =\\
    \sum_{I\cup I'=\{1,\ldots,n\}} j(I) P_{I,I'} j(I')
\end{equation}
\end{subequations}
instead of vertices as in~\eqref{eq:Comix}. It was pioneered by
\textsc{Comix} \cite{Gleisberg:2008fv} and
\textsc{OpenLoops}~\cite{Cascioli:2011va,Buccioni:2019sur}

Algebraically, all expressions~\eqref{eq:M} will give
the same final result, but the number of nodes that need to be evaluated
can vary slightly and numerical results will differ due to
the different order of evaluation.  \textsc{O'Mega}~\cite{Moretti:2001zz,Kilian:2007gr}
allows to compute the amplitude both as~\eqref{eq:Helac} and
as~\eqref{eq:Keystones} and confirms these expectations.

While it is impossible to give general estimates for the number of
wavefunctions that need to be evaluated in realistic models, one can count
them for some examples using \textsc{O'Mega}. In the standard model, it appears
that~\eqref{eq:Helac} requires some 10\%{} fewer evaluations
than~\eqref{eq:Keystones} in an optimal implementation.
One advantage of~\eqref{eq:Keystones} and~\eqref{eq:Comix} is that at
most~$n/2$ of the external momenta appear in the~$\phi$ compared
to~$n-1$ for~\eqref{eq:Helac}.  Therefore fewer steps with
accumulating floating point errors are required in
the recursive evaluation of~$\phi(I)$.
While this could in principle be a significant advantage in
amplitudes with strong gauge cancellations, the difference appears to
be small in practice.

The algorithm adding quantum numbers to~\eqref{eq:off-shell-nodes},
\eqref{eq:recursion} and~\eqref{eq:M} described in the following
sections is equally applicable for all three variants
in~\eqref{eq:M}.

\subsection{Evaluation}
\label{sec:evaluation}

In the case of a fixed physics model with a moderate number of fields
and couplings, such as the standard model, the recursive
evaluation~\eqref{eq:recursion} can be expressed as an iteration of
matrix multiplications~\cite{Caravaglios:1995cd,%
  Caravaglios:1996nq,Caravaglios:1998yr,Kanaki:2000ey,%
  Cafarella:2007pc,Gleisberg:2008fv,Actis:2012qn,Actis:2016mpe}.
This approach has the advantage that all scattering amplitudes in the
supported model can be computed using the same executable, without the
need for recompilation.

However, extending this approach to more complicated models,
in particular to models that can be specified by endusers
in formats like UFO~\cite{Degrande:2011ua,Darme:2023jdn}, is far from
trivial~\cite{Denner:2017wsf}.  Instead, it is beneficial to represent
the recursion relations~\eqref{eq:recursion} abstractly by a data
structure from which dedicated code can be generated and compiled subsequently,
following the pioneering treatment of Feynman diagrams in
\textsc{Madgraph}~\cite{Stelzer:1994ta,Alwall:2011uj}.  This approach has been
implemented for the recursive evaluation
in~\cite{Moretti:2001zz,Kilian:2007gr,Cascioli:2011va,Buccioni:2019sur}.

This motivates the search for a data structure that represents the
recursion relations~\eqref{eq:recursion} concisely and can be
constructed efficiently from the Feynman rules of a model.  The
obvious candidate is a finite Directed Acyclical Graph~(DAG), that
corresponds to the evaluation of an arithmetical expression in which
common subexpressions are evaluated only once and later recalled from
memory when needed again.

Additional benefits of algebraic manipulations are that it is easier
to prune the computation of wavefunctions that are not needed in the
final result, that one can target special hardware or dedicated
virtual machines~\cite{ChokoufeNejad:2014skp} that avoid the need for
compilation.  Formfactors can be restricted to lightlike momenta at
compile time~\cite{Alboteanu:2008my,Kilian:2014zja,Kilian:2015opv}.
Finally, one can optionally instrument the code with
numerical checks of Ward and Slav\-nov-Taylor identities for gauge boson
wavefunctions, in order to test matrix element
generator, numerical libraries and model descriptions.

\section{DAGs and Bundles}
\label{sec:DAG}

In this section, I will focus on universal mathematical
constructions, not practical algorithms.  The discussion of
the latter will follow in section~\ref{sec:feynman}.

Given a set~$N$ of \emph{nodes}, a set~$E$ of \emph{edges} and a
set~$C(N) \subseteq 2^N$ of \emph{children}
which is typically the set of subsets of nodes with a limited number
of elements,
any map from~$N$ to the powerset of~$E \times C(N)$
\begin{equation}
\label{eq:Delta}
  \Delta : N \to 2^{E \times C(N)}
\end{equation}
defines a Directed Graph~$\mathbf{G}=(N,E,\Delta)$ in the sense
described below.
The function~$\Delta$ can be specified completely by the
set~$\Set{(n,\Delta(n))|n\in N}$ of ordered pairs.  This equivalence
will be used below to define transformations on DAGs as set
theoretical operations that can be implemented efficiently in
computer programs. I will often
employ the more intuitive notation~$\Set{n\mapsto\Delta(n)|n\in N}$ or
the abbreviated form~$\Set{\delta_n|n\in N}$.
In order to avoid excessive nested superscripts, I will sometimes use the
notation~$A\to B$ for set~$B^A$ of all functions from the set~$A$ to the
set~$B$.

With wavefunctions as nodes and vertex factors as edges, this
definition captures the recursion relations~\eqref{eq:recursion}
exactly.  Note that the map~$\Delta$~\eqref{eq:Delta} is well defined
iff the combination of momenta and other quantum numbers identifies
the wavefunctions or currents uniquely.

There are cases where physical quantum numbers are not
sufficient to distinguish wavefunctions.  For example, if the
scattering amplitude is to be expanded in the powers of some coupling
constants, these powers can contribute at different levels of the
recursive expansion.  Therefore a wavefunction can appear more than
once with the same momentum and physical quantum numbers.  This forces
us to add the powers of these coupling constants as unphysical labels
that will be combined in the final step~\eqref{eq:M}.  Since such
labels are later required anyway to disambiguate
variable names in the generated numerical code, this adds no
additional burden.  Such a counting of coupling constants is of course
crucial for adding a consistent number of counterterms in calculations
involving loops and when adding precomputed loops using a skeleton
expansion or effective actions.

The nodes in the preimage of~$\emptyset$ under~$\Delta$
\begin{equation}
  L = \Delta^{-1}(\emptyset)
    = \Set{ n \in N | \Delta(n)=\emptyset }
\end{equation}
are called \emph{leaf nodes} and correspond to the external states in
scattering amplitudes.
Since the elements of~$C$ are sets of
elements of~$N$, we can derive from~$\Delta$ two mutually recursive
expansion functions
\begin{subequations}
\label{eq:Delta-hat-*}
\begin{equation}
  \label{eq:Delta-hat}
  \begin{aligned}
    \hat\Delta: E\times C(N) &\to E\times C(2^{E\times C(N)}) \\
     (e,\set{n_i|i\in I}) &\mapsto (e,\set{\Delta^*(n_1)|i\in I})
  \end{aligned}
\end{equation}
element-by-element and
\begin{equation}
  \label{eq:Delta*}
  \Delta^*(n) = \begin{cases}
     \{n\} & \text{for}\; \Delta(n)=\emptyset \\
     \hat\Delta(\Delta(n)) & \text{for}\; \Delta(n)\not=\emptyset\,.
  \end{cases}
\end{equation}
\end{subequations}
If~$\mathbf{D}=(N,E,\Delta)$ represents an acyclical graph, i.e.~a DAG,
with a finite number of nodes~$|N|$,
the functions~$\Delta^*$ and~$\hat\Delta$ will reach a fixed point
after a finite number of steps.  This fixed point
consists exclusively of mutually nested sets of leaves.
If the image of~$\Delta$ consists only of singleton sets
and~$\emptyset$, the fixed point reached from any starting node~$n$
corresponds to a tree diagram.  Otherwise it corresponds to a forest
of tree diagrams, if the elements of the sets are distributed recursively.

As an illustration, consider the DAG~$\mathbf{D}$ with the sets
\begin{subequations}
\label{eq:example}
  \begin{align}
    N &= \{1,2,\ldots,8\} \\
    E &= \emptyset \\
    C &= \Set{\{n,n'\}| n\not=n'\in N}
  \end{align}
and the map
\begin{multline}
  \Delta = \bigl\{ 1\mapsto\emptyset,\; 2\mapsto\emptyset,\;
     3\mapsto\emptyset,\; 4\mapsto\emptyset, \\
     5\mapsto\{\{1,2\}\},\; 6\mapsto\{\{5,3\}\},\; 7\mapsto\{\{5,4\}\}, \\
     8\mapsto\{\{6,4\}, \{7,3\}\} \bigr\}\,,
\end{multline}
where I have not spelled out the unlabeled edges.
\end{subequations}
A quick calculation gives
\begin{multline}
  \Delta^*(8) =
    \{\{\{\{\{\{1,2\}\}, 3\}\}, 4\},\\
      \{\{\{\{\{1,2\}\}, 4\}\}, 3\} \}\,.
\end{multline}
This corresponds to the forest consisting of the trees
\begin{subequations}
  \begin{align}
       &\{\{\{1,2\},3\},4\} \\
       &\{\{\{1,2\},4\},3\}\,.
  \end{align}
\end{subequations}

This DAG encodes a stripped down version of the Feynman diagrams for the
process~$e^+e^-q\bar q\to g$, 
that ignores both the details of the couplings and
the contributions of~$Z$ and Higgs bosons.  Note that the common
subdiagram~$e^+e^-\to\gamma$ appears only once
in the DAG as~$5\mapsto\{\{1,2\}\}$, but twice in the forest. 

A general directed graph can contain cycles and 
the functions~$\Delta^*$ and~$\hat\Delta$ will not reach a fixed point
even if~$|N|<\infty$. As described in section~\ref{sec:momenta}, it
will however always
be possible to equip~$N$ with a natural order
so that the application of~$\Delta$ acts strictly decreasing with
respect to this order.
There can obviously be no cycles and the graph is guaranteed to
be a DAG in this case.

If the same node~$n$ appears many times in the children,
a DAG provides a very efficient encoding of large
sets of graphs.  The storage and computing time required by typical
sets of tree diagrams
grows factorially with the number of leaves~$|L|$.  In contrast, the
space and time required for implementing the DAG scales linearly with~$|N|$,
which only grows as an exponentially in~$|L|$.
Using persistent functional data
structures~\cite{Okasaki/PFDS}
instead of mutable arrays to
implement the function~$\Delta$ simplifies the algorithm described
below significantly. The additional space and time
requirements replace~$|N|$ by~$|N|\ln|N|$ and turn out not to be
important for large~$|N|$.

\subsection{Constructing DAGs}
\label{sec:constructors}

Using DAGs as a compact representation has only a marginal benefit if
their construction requires the generation of all tree diagrams in
intermediate steps or if the applications require a full expansion.
Fortunately, the sum of Feynman diagrams encoded in the DAG can be
evaluated either using the DAG directly or by generating
a dedicated numerical code that
evaluates each node~$n\in N$ only once.  As explained in
section~\ref{sec:feynman}, it turns out that the DAGs
representing perturbative scattering amplitudes can be constructed
without requiring the construction of the corresponding
forest.

For this purpose, I introduce the empty DAG
\begin{equation}
\label{eq:epsilon}
  \epsilon = (\emptyset, \emptyset, \emptyset, \emptyset)\,
\end{equation}
where~$\Delta=\emptyset$ is the function with empty domain and
codomain.  I also define a function
\begin{subequations}
\label{eq:omega}
\begin{equation}
  \begin{aligned}
    \omega : (N \to E \times 2^{C(N)}) \times \mathcal{D} &\to \mathcal{D}\\
        (n\mapsto (e, c),x) &\mapsto \omega_{n\mapsto (e, c)}(\mathbf{D})
  \end{aligned}
\end{equation}
with the function~$\omega_{n\mapsto (e, c)}$ that
adds a node~$n$ together with the mapping~$n\mapsto(e,c)$
\begin{multline}
  \omega_{n\mapsto (e, c)} (N, E, \Delta) = \\
    \left( N \cup \{n\}, E \cup e, \Delta \cup\{n\mapsto(e,c)\} \right)\,,
\end{multline}
\end{subequations}
where $e$ and~$(e,c)$ are shorthands for the sets~$\Set{e_i|i\in I}$
and~$\Set{(e_i,c_i)|i\in I}$ with $|I|$~elements.  In particular, they
may be empty to allow inserting a leaf node. In order to avoid
ambiguities in the definition of~$\omega$, I will
require that~$n\not\in N\land n'_i\in N$
in~$\omega_{n\mapsto\{(e,\set{n'_i|i\in I})\}}$.

With these definitions, the DAG in~\eqref{eq:example} is
\begin{multline}
  \omega_{8\mapsto\{\{6,4\},\{7,3\}\}}
  \omega_{7\mapsto\{\{5,8\}\}} \omega_{6\mapsto\{\{5,3\}\}}\\
  \omega_{5\mapsto\{\{1,2\}\}} \omega_{4\mapsto\emptyset}
  \omega_{3\mapsto\emptyset}\omega_{2\mapsto\emptyset}\omega_{1\mapsto\emptyset}\epsilon\,,
\end{multline}
where the function applications associate to the right, of course.  It
is obvious that any finite DAG can be
constructed by repeated applications of~$\omega$.

For the finite DAGs that are the subject of this paper,
the function~$\omega$ can be implemented easily in
programming languages that have efficient support for persistent sets
and maps (also known as dictionaries) that can grow without a lot of
reallocation.  Functional programming languages with
garbage collection make such implementations particularly
straightforward.
The domain and codomain of functions like~$\omega$~\eqref{eq:omega}
are highly structured sets and static type systems allow to verify
already at compile time
that only matching functions are being composed.  Beyond
preventing errors, a strict type discipline helps to uncover
mathematical structures, such as the ones described in this section.
This paper is based on the implementation
in the matrix element generator
\textsc{O'Mega}~\cite{Moretti:2001zz} using
\texttt{ocaml}~\cite{ocaml5/manual}, as
described in appendix~\ref{app:DAG-ocaml}.

\subsection{Lattices of DAGs}
\label{sec:lattice}

For our purposes, DAGs representing scattering amplitudes for the same
external states, categories of DAGs that share the same leaf nodes
\begin{equation}
  \mathcal{D}_L =
     \Set{ \mathbf{D} = (N, E, \Delta) | \Delta^{-1}(\emptyset) = L }
\end{equation}
are the most interesting.
Since we describe a DAG as a tuple of sets, there is a natural notion of
inclusion for pairs of DAGs in~$\mathcal{D}_L$
\begin{subequations}
\label{eq:inclusion}
\begin{multline}
  \mathbf{D}' = (N',E',\Delta') \subseteq \mathbf{D} = (N,E,\Delta) \Leftrightarrow \\
    N' \subseteq N \land E' \subseteq E 
    \land \left( \forall n \in N': \Delta'(n) \subseteq \Delta(n) \right)\,.
\end{multline}
\end{subequations}
It is obvious that this notion of inclusion corresponds to the inclusion
of the sets of tree diagrams encoded by the DAGs.

In the same fashion, we can define union and intersection for the
DAGs~$\mathbf{D}_i = (N_i,E_i,\Delta_i)$
\begin{subequations}
\begin{align}
  \mathbf{D}_1 \cup \mathbf{D}_2 &= (N_1 \cup N_2, E_1 \cup E_2, \Delta_1 \cup \Delta_2) \\
  \mathbf{D}_1 \cap \mathbf{D}_2 &= (N_1 \cap N_2, E_1 \cap E_2, \Delta_1 \cap \Delta_2)
\end{align}
where 
\begin{multline}
\label{eq:union}
  \Delta_1 \cup \Delta_2 = \\
   \Set{ n \mapsto \Delta_1(n) \cup \Delta_2(n) | n\in N_1\cap N_2} \\
   \cup \Set{ n \mapsto \Delta_1(n) | n\in N_1\setminus N_2} \\
   \cup \Set{ n \mapsto \Delta_2(n) | n\in N_2\setminus N_1}
\end{multline}
and in
\begin{multline}
\label{eq:intersection}
  \Delta_1 \cap \Delta_2 = \\
  \bigl\{ n \mapsto \Delta_1(n) \cap \Delta_2(n) \bigr| n\in N_1\cap N_2 \\
    \land \left( \Delta_1(n) \cap \Delta_2(n) \not= \emptyset \lor n \in L \right) \bigr\}
\end{multline}
I am careful to avoid adding new leaf nodes to the intersection.
\end{subequations}

From these definitions, it is obvious that~$\subseteq$
turns~$\mathcal{D}_L$ into a \emph{partially ordered set} and~$\cup$
and~$\cap$ turn it into a \emph{lattice}.
From this point of view,
$\mathbf{D}_1\cup\mathbf{D}_2$ is the least common upper bound
of~$\mathbf{D}_1$ and~$\mathbf{D}_2$,
while~$\mathbf{D}_1\cap\mathbf{D}_2$ is their greatest common lower
bound.  Finally~$\mathcal{D}_L$ is bounded from below, with
\begin{equation}
  \bot_L = (L,E,\set{n\to\emptyset|n\in L})
\end{equation}
as the bottom element.

\subsection{Mapping and Folding DAGs}
\label{sec:folds}

The most important functions for manipulating DAGs and extracting the
information encoded by them are
\emph{folds} that perform a nested application of a suitable function
for all nodes to a starting value~$x$
\begin{equation}
\label{eq:Phi}
  \Phi_f ((N, E, \Delta), x) = f_{\delta_{|N|}} \cdots f_{\delta_2}f_{\delta_1}x\,,
\end{equation}
where the elements
of~$\Delta=\{\delta_{n_1},\delta_{n_2},\ldots,\delta_{n_{|N|}}\}$ are
arranged in the partial order if the nodes that guarantees acyclicity of the DAG.
The only constraint on the function
\begin{equation}
  \begin{aligned}
    f : (N \to E \times 2^{C(N)}) \times X &\to X\\
        (\delta,x) &\mapsto f_{\delta}(x)
  \end{aligned}
\end{equation}
is that the domain and codomain of~$f_{\delta}:X\to X$ must be identical.
The computational cost scales with
the size of the DAG and not with the size of the forest of tree
diagrams described by it.

Used with the constructor~$\omega$~\eqref{eq:omega}
on the empty DAG, the fold
performs a complete copy of any DAG~$\mathbf{D}$
\begin{equation}
\label{eq:copy-of-fold}
  \Phi_\omega(\mathbf{D},\epsilon) = \mathbf{D}\,.
\end{equation}

Precomposing the first argument of~$\omega$ in~\eqref{eq:copy-of-fold} with
a function
\begin{equation}
    f : (N\to 2^{E\times C}) \to (N\to 2^{E\times C})
\end{equation}
in the first argument using the notation
\begin{equation}
\label{eq:precomposition}
  (\omega\circ f)_{\delta} = \lambda_{f(\delta)}
\end{equation}
maps a DAG~$\mathbf{D}$ to a new DAG~$\mathbf{D}'$
\begin{equation}
\label{eq:map-of-fold}
  \Phi_{\omega\circ f}(\mathbf{D},\epsilon)
     = \mathbf{D}'
\end{equation}
which can encode a different set of tree graphs.

The precomposition~\eqref{eq:precomposition} can naturally be
extended to functions mapping nodes to sets of nodes
\begin{equation}
  \begin{aligned}
    f : (N\to 2^{E\times C}) &\to 2^{(N\to 2^{E\times C})} \\
        \delta &\mapsto \{f_1(\delta),\ldots,f_k(\delta)\}
  \end{aligned}
\end{equation}
as
\begin{equation}
\label{eq:precomposition1}
  \omega_{f(\delta)} = \omega_{f_k(\delta)}\ldots\omega_{f_1(\delta)}
\end{equation}
with the identity
\begin{equation}
\label{eq:precomposition0}
    \omega_{\emptyset} \mathbf{D} = \mathbf{D}
\end{equation}
iff the result of~$f$ is the empty set~$\emptyset$.

Finally, I define a function
\begin{equation}
\label{eq:harvest}
  H : (S, \mathbf{D}) \mapsto \mathbf{D}' \subseteq \mathbf{D}
\end{equation}
that takes a set~$S\subseteq N$ of nodes and a DAG and returns the minimal DAG that
contains all the nodes in the set such that the mutually recursive
evaluation of the functions~$\Delta^*$
and~$\hat\Delta$ from~\eqref{eq:Delta-hat-*} is well defined for the
nodes in this set.  Intuitively, this corresponds to following all
chains of arrows in~$\Set{n\to\Delta(n)|n\in N}$ from~$\mathbf{D}$ that start in~$S$.

\subsection{Bundles}
\label{sec:bundles}

I am interested in maps between DAGs that respect certain structures. In
order to describe these concisely, I borrow the notion of bundles
from topology and differential geometry.

A \emph{bundle}~$\mathbf{B} = (X, B, \pi)$ is a triple consisting of
a set~$X$, called the \emph{total set}, a set~$B$, called the \emph{base}, and a
\emph{projection}~$\pi:X\to B$.  
The preimages~$\pi^{-1}(b)\subseteq X$ are
called \emph{fibers}. The notation~$\pi^{-1}:B\to 2^X$ must of course
not be misunderstood as the inverse of~$\pi$. The fibers are pairwise
disjoint and their union
\begin{equation}
  X = \bigsqcup_{b\in B} \pi^{-1}(b)
\end{equation}
reproduces the set~$X$.  A \emph{section} is a map~$s:B\to X$ for
which~$\pi\circ s: B\to B$ is the identity.  It corresponds to
choosing one and only one element from each fiber.
This definition generalizes the trivial bundle
\begin{subequations}
\begin{equation}
  \mathbf{B}_{\text{trivial}} = (B\times F, B, \pi)
\end{equation}
with
\begin{align}
  \pi(b, x) &= b \\
  \pi^{-1}(b) &= (b, F)
\end{align}
where all fibers are trivially isomorphic to~$F$ and a section is the
parameterized graph~$s:B\to B\times F$ of a function~$B\to F$.
\end{subequations}

Bundles formalize equivalence relations on the set~$X$,
with the base~$B$ as the set of all equivalence classes and~$\pi$ the
canonical projection of an element of~$X$ to its equivalence class.
The composition~$\pi^{-1}\circ\pi:X\to2^X$ maps each element to the
set of the members of its equivalence class.  Sections
correspond to choosing one element from each equivalence class. An
illustrative example is equivalence of nodes up to color quantum
numbers, where~$\pi$ corresponds to ignoring color. Flavor,
coupling constant and loop expansion order can be treated in the same
way.

Bundles can be arranged in a sequence
\begin{equation}
\label{eq:B-complex}
  \begin{CD}
    B_0           @<\pi_1<<          B_1           @<\pi_2<<         B_2            @<\pi_3<< \cdots\,.
  \end{CD}
\end{equation}
However, since the preimage~$\pi_i^{-1}$ is not the inverse of the
projection~$\pi_i$, the preimage of a composition of projections is not the
composition of the individual preimages, but
\begin{equation}
  (\pi_{i}\circ\pi_{i+1})^{-1}(b) = \cup_{x\in\pi_{i}^{-1}(b)} \pi_{i+1}^{-1}(x)
\end{equation}
instead.

As in the case of DAGs, such structures and the operations on them
can be implemented for finite sets~$X$
straightforwardly in functional programming languages with static type
systems and garbage collection (cf.~appendix~\ref{app:bundle-ocaml}).
In particular, it is efficient to add
elements to the set~$X$ and update the base~$B$ and maps~$\pi$
and~$\pi^{-1}$ immediately.  This allows to grow a bundle
simultaneously while
building a new DAG in order to maintain the relationships to be introduced
in section~\ref{sec:projections}.

\subsection{Projections and Preimages of DAGs}
\label{sec:projections}

Given a DAG~$\mathbf{D}=(N,E,\Delta)$, where the set of
nodes~$N$ is also the total set in a bundle~$\mathbf{B}=(N,B,\pi)$, it is
natural to ask if there is a canonical
DAG~$\mathbf{D}'=(B,E',\Delta')$
with the base of~$\mathbf{B}$ as its set of nodes.

First, we observe that every section~$s$ of~$\mathbf{B}$ and
map~$f:E\to E'$ defines a projected
DAG~$\mathbf{D}_{s,f}=(B,E',\Delta_{s,f})$ with
\begin{subequations}
\label{eq:DAG-projection}
\begin{equation}
  \begin{aligned}
    \Delta_{s,f} : B&\to 2^{E' \times C(B)} \\
                  b&\mapsto \hat\pi_f(\Delta(s(b)))
  \end{aligned}
\end{equation}
where~$\hat\pi_f$ is the distribution of~$\pi$ over the nodes together
with the application of~$f$ to the edges
\begin{equation}
   \hat\pi_f(e,\set{n_i|i\in I}) = (f(e),\set{\pi(n_i)|i\in I})\,.
\end{equation}
\end{subequations}
The formula~\eqref{eq:DAG-projection} has to be augmented by the
prescription that a~$b$ for which~$s(b)$ is a leaf node
in~$\mathbf{D}$ and therefore~$\Delta_{s,f}(b)=\emptyset$
is \emph{not} added as a leaf node to~$\mathbf{D}_{s,f}$, similar to
the definition~\eqref{eq:intersection} of the intersection of two DAGs.

In most
cases~$f:E\to E'$ will be a simple projection that in our applications
will be determined
straightforwardly by the two sets of Feynman rules governing the
construction of the two DAGs.  Therefore we can
write~$\mathbf{D}_{s}$ instead of the more
explicit~$\mathbf{D}_{s,f}$.

The dependence of this projection on the section~$s$ is not
satisfactory.  However, the DAG
\begin{equation}
  \Pi(\mathbf{D}) = \bigcup_{s\in S(\mathbf{B})} \mathbf{D}_s\,,
\end{equation}
where~$S(\mathbf{B})$ denotes the set of all sections of the bundle~$\mathbf{B}$,
is well defined and will be shown to suit our needs.
Observe that the union is
the correct universal construction for our applications, because the
additional quantum numbers in~$N$ lead to more selection rules.
These selection rules are the reason for the dependency of~$\mathbf{D}_s$ on~$s$.
The DAG corresponding
to the more basic set of nodes~$B$ should therefore be the combination of all
possibilities.
As an example consider the scattering of two scalars without and with
flavor. Without flavor, there will be $s$-, $t-$ and $u$-channel
diagrams.  With a conserved flavor, only one of them will remain.

Note however, that this construction does \emph{not} guarantee
that the set of nodes of the DAG~$\Pi(\mathbf{D})$ is actually the
full base~$B$
of the bundle~$\mathbf{B}$.  We must therefore demand in addition compatibility of
DAG and bundle, by requiring that the diagram
\begin{equation}
\label{eq:B-D-compatibility}
  \begin{CD}
    B               @<\pi<<  N \\
    @A\nu AA                 @AA\nu A \\
    \Pi(\mathbf{D}) @<\Pi<<  \mathbf{D}
  \end{CD}
\end{equation}
commutes.  The function~$\nu$ in the commuting square~\eqref{eq:B-D-compatibility} just
extracts the set of nodes from a DAG
\begin{equation}
  \nu(N,E,\Delta) = N\,.
\end{equation}
The objects in the commuting square~\eqref{eq:B-D-compatibility} can
be understood as a combination of a pair of DAGs and a bundle, which I
will call a \emph{fibered DAG}. In programs, nodes can be added
to the DAG~$\mathbf{D}$ and the bundle in concert such that the
relationship~\eqref{eq:B-D-compatibility} is maintained.

An immediate benefit of such an universal construction of the projection is
that it provides a corresponding preimage~$\Pi^{-1}$ which maps DAGs with
the base~$B$ as nodes to all DAGs with the set~$N$ as nodes.
The maps in the preimage can be written
\begin{equation}
\label{eq:DAG-lift}
  \begin{aligned}
    \Delta^{s,f} : N&\to 2^{E \times C(N)} \\
                  n&\mapsto \hat s^f(\Delta(\pi(n)))\,
  \end{aligned}
\end{equation}
where~$\hat s^f$ is to be understood as the distribution of~$s$ over
the nodes together with the application of~$f$ to the edges.  Unfortunately,
in contrast to~\eqref{eq:DAG-projection}, there will not be a single
function~$f:E\to E'$.  Instead, we must allow that~$\hat s^f$ maps
into the powerset~$2^{E'\times C(N)}$ instead of~$E'\times C(N)$.  In
addition, the image of~$f$ will depend, via the Feynman rules, on the
nodes appearing as children.

Since the resulting notation would be unnecessarily cumbersome, I will
refrain from making the nature of~$f$ in~\eqref{eq:DAG-lift} explicit as
a function by specifying its domain and writing out all of its
arguments.  Nevertheless, the discussion of the example in
section~\ref{sec:algorithm} will demonstrate how a set of Feynman
rules defines the maps~$\Delta^{s,f}$ unambiguously.

In this picture, the application of Feynman rules amounts to choosing
a particular element of the preimage~$\Pi^{-1}$.  It would however be extremely
wasteful to construct the preimage first and to throw away all but one
of its elements later.  In section~\ref{sec:algorithm}, I will describe an
algorithm that can be used to construct the desired element directly.

So far, I have assumed that the DAGs are selected by Feynman
rules that are local to each vertex in the case of Feynman diagrams or
to each element~$\delta_i$ of the map~$\Delta$ in our DAGs
individually.  There are however important exceptions.  The most
important is provided by loop expansions.  There it is required for
consistency that counter\-terms are inserted a fixed number of times in
Feynman diagrams.  Such conditions on complete Feynman diagrams do
not translate immediately to the DAGs, whose components can enter the
scattering amplitudes~\eqref{eq:M} more than once.  Fortunately, this
problem can be solved by introducing additional unphysical labels
representing loop orders to the physical labels of the nodes and to
select the required combinations of wavefunctions in~\eqref{eq:M} at
the end, as will be described in section~\ref{sec:orders}.
The same applies to selecting fixed orders in the perturbative
expansions, as required for comparing to many results from the literature.

We call two DAGs~$\mathbf{D}_1=(N_1,E_1,\Delta_1)$
and~$\mathbf{D}_2=(N_2,E_2,\Delta_2)$ 
equivalent with respect to a
pair of bundles~$\mathbf{B}_1=(N_1,B,\pi_1)$
and~$\mathbf{B}_2=(N_2,B,\pi_2)$ with the same base~$B$ iff there is a
common projected DAG~$\mathbf{D}$
\begin{equation}
  \Pi_1(\mathbf{D}_1) = \mathbf{D} = \Pi_2(\mathbf{D}_2)\,.
\end{equation}
In this case~$\mathbf{D}_1$ and~$\mathbf{D}_2$ can be viewed as
refinements of the same basic DAG~$\mathbf{D}$.
This notion of equivalence generalizes the notion of \emph{topological
equivalence} for diagrams, where two diagrams are considered
equivalent if they agree after stripping off all quantum numbers.
With the new notion of equivalence, we can say that the sets of
Feynman diagrams encoded in a DAG are equivalent up to flavor or upto
color. 

Using the basic commuting square~\eqref{eq:B-D-compatibility}, we can
immediately extend the bundle complex~\eqref{eq:B-complex} to include the
corresponding DAGs
\begin{equation}
\label{eq:B-D-complex}
  \begin{CD}
    B_0          @<\pi_1<< B_1          @<\pi_2<< B_2          @<\pi_3<< \cdots \\
    @A\nu AA               @A\nu AA              @A\nu AA                      \\
    \mathbf{D}_0 @<\Pi_1<< \mathbf{D}_1 @<\Pi_2<< \mathbf{D}_2 @<\Pi_3<< \cdots
  \end{CD}\,.
\end{equation}
In our applications, this complex does not continue further to the left,
because for each number of leaf nodes there is a natural leftmost nontrivial
DAG~$\mathbf{D}_{\mathrm{P}}$, described in section~\ref{sec:momenta}
below.

In the following section~\ref{sec:feynman} I will describe how to use
Feynman rules to walk the lower row of~\eqref{eq:B-D-complex} to
construct a DAG for a scattering amplitude efficiently in stages.

\section{DAGs from Feynman Rules}
\label{sec:feynman}

In principle, it is possible to construct the DAG encoding all Feynman
diagrams in a single step.

First one adds leaf nodes for external states,
labeled by all quantum numbers (momentum, spin/polarization, flavor,
color, \ldots).  Which states are to be included here depends on the
choice of algorithm, as has been discussed in
section~\ref{sec:topologies}.

Then one uses the Feynman rules of the model to add all nodes where the
node and its children correspond to an allowed vertex.  This proceeds
iteratively: in the first step all subsets of the leaf nodes appear as
children.  In the following steps subsets of all nodes, including the
leaf nodes appear as children subject to the constraint that no leaf
node appears twice if the DAG is expanded recursively with the
functions~$\Delta^*$ and~$\hat\Delta$ from~\eqref{eq:Delta-hat-*}.
This iteration will
terminate after a finite number of steps when all leaf nodes have been
combined in all possible ways.
While this algorithm inserts nodes that will not appear in
the scattering amplitude the function~$H$~\eqref{eq:harvest} can
be used to harvest the minimal DAG.

This is a workable approach, but it is neither the most efficient
nor particularly maintainable in actual code.  Since the nodes are labeled by all
quantum numbers, handling them all at once requires the construction
of many nodes that will not appear in the final result. Adding
quantum numbers in several stages instead allows us to use the constraints from
earlier simpler stages to avoid in later stages the construction of
many more complicated nodes that will never be used. While not
relevant for the final numerical code, experience with early versions
of \textsc{O'Mega}~\cite{Moretti:2001zz,Kilian:2007gr} revealed that the
latter approach requires noticeably less time and memory for constructing the code.

Breaking up the
construction of the DAG into several stages also simplifies the implementation of
each stage and allows separate testing and swapping of different
implementations.  Finally, applications often need access to projected
DAGs as described in section~\ref{sec:projections} anyway.  A
prominent example is the construction of phase space parameterizations
that only refer to kinematical information, such as propagators and masses.

Some of the stages described in the following subsections will be
performed in a particular order, while the order of others can be
interchanged easily.

\subsection{Momenta}
\label{sec:momenta}

An element of the set~$N_{\mathrm{P}}$ of nodes in the first
DAG~$\mathbf{D}_{\mathrm{P}}=(N_{\mathrm{P}},\emptyset,\Delta_{\mathrm{P}})$ to be constructed 
is uniquely labeled by
a subset of the powerset~$2^{\{1,2,\ldots,n\}}$ of labels for the
external momenta and the edges are unlabeled.  The leaf
nodes are the elements~$n(\{i\})$ of~$N_{\mathrm{P}}$ and the action of the
map~$\Delta_{\mathrm{P}}$ is given by
\begin{multline}
\label{eq:topology}
  n(I) \mapsto
  \Bigl\{ (\emptyset, \set{n(I_i)|1\le i\le k}) \Bigr| 2\le k\le l-1 \\
          \land \cup_{i=1}^kI_i=I \land I_i\not=\emptyset \Bigr\}
\end{multline}
where~$l$ is the maximum number of legs of the vertices in the model.
Obviously, we can order the nodes~$n(I)$ according to the number of
elements of~$I$ to prove that there are no cycles in~$\mathbf{D}_{\mathrm{P}}$.

In case of~\eqref{eq:Helac}, we only need the elements
of~$2^{\{2,\ldots,n\}}$ as labels. In the cases~\eqref{eq:Keystones}
and~\eqref{eq:Comix}, only labels with at most $n/2$~elements are needed.
Finally, the function~$H$~\eqref{eq:harvest} is applied to construct the
minimal DAG required for evaluating one of the expressions~\eqref{eq:M}.

\subsection{Flavors and Lorentz Structures}
\label{sec:flavor}
\label{sec:algorithm}

In the next stage, the momenta of the leaf nodes of~$\mathbf{D}_{\mathrm{P}}$ are
combined with the flavor quantum numbers of the corresponding
external state.  The resulting leaf nodes form
the starting point of a new
DAG~$\mathbf{D}_{\mathrm{F}}=(N_{\mathrm{F}},V_{\mathrm{F}},\Delta_{\mathrm{F}})$ and
bundle~$\mathbf{B}_{\mathrm{F}}=(N_{\mathrm{F}}, N_{\mathrm{P}}, \pi_{\mathrm{F}})$.
The edges~$V_{\mathrm{F}}$ are vertex factors consisting of coupling constants,
Lorentz tensors and Dirac matrices.

Using a fold~$\Phi$ of~$\mathbf{D}_{\mathrm{P}}$ using the constructor~$\omega$
of~$\mathbf{D}_{\mathrm{F}}$ with
precomposition~\eqref{eq:map-of-fold} that
maintains the fibration~\eqref{eq:B-D-compatibility} will ensure that
the nodes
of~$\mathbf{D}_{\mathrm{P}}$ are visited in the correct order of growing label
sets.  The function~$f$ that is precomposed to~$\omega$
in~\eqref{eq:map-of-fold} acts on each element
\begin{equation}
  n(I) \mapsto (\emptyset, \set{n(I_i)|1\le i\le k})
\end{equation}
of the map~$\Delta_{\mathrm{P}}$ as follows:
since the~$n(I_i)\in N_{\mathrm{P}}$ have been
processed, they are elements of the base of the growing
bundle~$\mathbf{B}_{\mathrm{F}}$.  Therefore, the fibers~$\pi_{\mathrm{F}}^{-1}(n(I_i))$ are
already complete and we can compute their cartesian product
\begin{equation}
  \Gamma = \pi_{\mathrm{F}}^{-1}(n(I_1)) \times \pi_{\mathrm{F}}^{-1}(n(I_2)) \times \cdots\,.
\end{equation}
We then use the Feynman rules to select all elements of~$\Gamma$ that can be
combined with another flavor to obtain a valid vertex.  This
defines a function~$\Gamma\mapsto2^{V_{\mathrm{F}}}$.
For each of the resulting
flavors, a new node labeled by~$I$ and this flavor is added
to~$\mathbf{D}_{\mathrm{F}}$ and~$\mathbf{B}_{\mathrm{F}}$
together with the corresponding vertex factors and
elements of~$\Gamma$ as edges and children, maintaining the
fibration~\eqref{eq:B-D-compatibility}.

This algorithm has been implemented in
\textsc{O'Mega}~\cite{Moretti:2001zz,Kilian:2007gr}
and is completely independent of the kind of Feynman
rules.  It can accommodate both hardcoded rules and rules derived from a
UFO file~\cite{Degrande:2011ua,Darme:2023jdn}.  The only potential performance
bottleneck is the efficient matching of vertices to the elements
of a~$\Gamma$ representing a large number of children.  For vertices with few legs, this is not a
practical issue, but care has to be taken for vertices with many legs
where the factorial growth of the number of permutations might be felt.

Once the flavors have been assigned, it is known which fermion lines
contribute in the computation of each node.  This information must
also be
added to the node in order to be able to assign the correct sign to
interfering contributions in~\eqref{eq:M} later.  Special care must be taken if the model
contains Majorana
fermions~\cite{Denner:1992me,Reuter:2002gn,Ohl:2002jp}.

By construction, after the fold is complete, the new
DAG~$\mathbf{D}_{\mathrm{F}}$ encodes all the information needed to
compute the scattering amplitude for the
leaf nodes in a theory without color, using one of the
formulae~\eqref{eq:M}.
Some nodes in~$\mathbf{D}_{\mathrm{F}}$ might not be needed due
to conserved quantum numbers.  Therefore the
function~$H$~\eqref{eq:harvest} from~$\mathbf{D}_{\mathrm{F}}$ is applied again to construct the
minimal DAG required to evaluate one of the expressions~\eqref{eq:M}.

\subsection{Colors}
Since the color representation depends on the flavor, the assignment
of color quantum numbers in the construction of the
DAG~$\mathbf{D}_{\mathrm{C}}=(N_{\mathrm{C}},V_{\mathrm{C}},\Delta_{\mathrm{C}})$
naturally comes after the construction of~$\mathbf{D}_{\mathrm{F}}$.

We can now follow the steps of the previous stage, as
described in section~\ref{sec:algorithm}, word for word, only
replacing the subscripts~$(\mathrm{F},\mathrm{P})$ by ~$(\mathrm{C},\mathrm{F})$.
The implementation in \textsc{O'Mega} uses the realization of the color flow
basis described in~\cite{Kilian:2012pz}, but, except for the labeling of
the nodes in~$N_{\mathrm{C}}$, the form of the vertices in~$V_{\mathrm{C}}$
and the Feynman rules to be used, the algorithm is
completely independent of the representation of the color algebra.

Having the color information available algebraically allows to compute
color factors and color correlators~\cite{Catani:1996vz} analytically.

\subsection{Coupling Orders}
\label{sec:orders}

As already mentioned in section~\ref{sec:projections}, there are cases
where it is important that the Feynman diagrams encoded by the DAG
contain certain coupling constants with fixed powers.  The most
important examples are the counterterms and the terms of an effective
action in a loop expansion.  Also the inclusion of self energy type terms
will not terminate in at DAG, unless a finite maximum expansion order
is prescribed.

For practical purposes it is sometimes also important to compute only
a part of a scattering amplitude corresponding to fixed powers of
couplings.  Such results are often available in the literature from Feynman
diagram based calculations and a comparison for the purpose of
validation is only possible if the DAG based calculation can select
exactly the same contributions.

A priori, this conflicts with the representations~\eqref{eq:M} of
scattering amplitudes as DAGs, since the wavefunctions or currents
will have accumulated different powers of couplings that will be mixed
by~\eqref{eq:M}.

Fortunately, there is a simple solution.  For example, in the case
of~\eqref{eq:Keystones} we can write
\begin{multline}
\label{eq:Keystones-orders}
  \mathcal{M}_o(\{1,2,\ldots,n\}) =\\
    \sum_{\substack{I_1\cup I_2\cup I_3=\{1,\ldots,n\}\\o_1+o_2+o_3=o}}
          K_{I_1,I_2,I_3} \phi_{o_1}(I_1)\phi_{o_1}(I_2)\phi_{o_1}(I_3)
\end{multline}
to compute the scattering amplitude at the coupling order~$o$.  The
only change required is that the wavefunctions have to keep track of
the coupling orders accumulated in their recursive computation.  Since
the powers of the couplings are additive, we never have to add the
wavefunctions or currents that exceed the requested order to the DAG.

This necessitates augmenting the set of labels of the nodes by unphysical
``quantum numbers'' corresponding to the coupling orders. It can be
implemented easily, as long as the number of coupling orders to be tracked
remains moderate.

\subsection{Skeleton Expansion}
\label{sec:skeleton}

If we are using DAGs to efficiently implement a skeleton expansion,
the remarks in section~\ref{sec:orders} apply word for word by
replacing ``coupling order'' by ``loop order''.

\subsection{Multiple Amplitudes}
\label{sec:multiple}

In practical applications~\cite{Kilian:2007gr}, it is usually necessary
to compute scattering amplitudes for the same external
momenta, but more than one combination
of flavors and colors at the same time.  These flavor and color
combinations often overlap pairwise and the sets of
leaf nodes will also overlap, i.e.~$L_1\cap L_2\not=\emptyset$.  In this
case, it is efficient to combine the corresponding
DAGs~$\mathbf{D}_{L_1}$ and~$\mathbf{D}_{L_2}$ into a single DAG and to compute the
scattering amplitudes from this DAG in order to reuse nodes
from the part of the DAG build on~$L_1\cap L_2$.  For this purpose,
we can generalize the
union defined in~\eqref{eq:union} to a map
\begin{equation}
  \cup: \mathcal{D}_{L_1} \times \mathcal{D}_{L_2} \to \mathcal{D}_{L_1\cup L_2}
\end{equation}
in an obvious way.

\section{Code Generation}
\label{sec:code}

The example~\eqref{eq:example} can be translated directly into, e.\,g.~Fortran, as
\begin{verbatim}
  w1 = phi(p1)
  w2 = phi(p2)
  w3 = phi(p3)
  w4 = phi(p4)
  p5 = p1 + p2
  w5 = prop(p5)*g*w1*w2
  p6 = p5 + p3
  w6 = prop(p6)*g*w5*w3
  p7 = p5 + p4
  w7 = prop(p7)*g*w5*w4
  p8 = p6 + p4
! p8 = p7 + p3
  w8 = prop(p7)*(g*w6*w4 + g*w7+w3)
\end{verbatim}
where~\verb+p+$n$ and~\verb+w+$n$ denote fourmomenta and
wavefunctions, respectively.  \verb+phi()+ computes external
wavefunctions, \verb+prop()+ propagators and \verb+g+ is a coupling
constant.  Using overloaded operators~\verb-+-,~\verb+-+ and~\verb+*+
allows to write similarly concise and readable code for realistic
models with standard model quantum numbers.
In the case of
more general models, functions implementing the vertex factors can
be generated from UFO files~\cite{Degrande:2011ua,Darme:2023jdn}.

Identically structured code can be
emitted as bytecode for a virtual machine that realizes the operators
as basic instructions~\cite{ChokoufeNejad:2014skp}.  The improving memory
bandwidth for graphical processing units even allows to start
targeting GPUs for interesting examples.

As already mentioned in the introduction, the generation of robust
numerical code is much more challenging if the DAG encodes diagrams
that contain loops.  The problem has been solved for the standard
model~\cite{Cascioli:2011va,Buccioni:2019sur,Actis:2012qn,Actis:2016mpe}.
The structures described in the paper will help with the task of
extending this approach to general models.

\section{Conclusions}
I have described the algebraic structures that organize recursive
calculations in perturbative quantum field theory without the need to
expand intermediate expressions into Feynman diagrams.  In functional
programming languages, these algebraic structures translate
directly into data structures.  In a second step, these data
structures are translated to efficient numerical code for any
programming language or hardware target required.

This algebraic approach adds flexibility over purely numeric
implementations tied to specific models and computing targets.
It allows for more extensive
consistency checks and paves the way for more challenging
applications.

\backmatter
\bmhead{Acknowledgments}
I thank Wolfgang Kilian, J\"urgen Reuter and the other members of the \textsc{Whizard}
team for the decades long productive collaboration.
This work is supported by the German Federal Ministry for Education and
Research (BMBF) under contract no.~05H21WWCAA.


\begin{appendices}
\section{Implementation}
\label{app:ocaml}

\subsection{DAGs}
\label{app:DAG-ocaml}
Here is the relevant subset of the \texttt{ocaml}
signature~\cite{ocaml5/manual} of the \texttt{DAG} module in
\textsc{O'Mega}~\cite{Moretti:2001zz}, implementing the functions from
sections~\ref{sec:constructors} and~\ref{sec:folds}.  For flexibility,
this module is implemented as a functor application on the
types~\verb|node|, \verb|edge| and \verb|children|, corresponding
to~$N$, $E$ and~$C(N)$ respectively
\begin{verbatim}
module type DAG = sig
  type node
  type edge
  type children
  type t
  val empty : t
  val add_node : node -> t -> t
  val add_offspring :
    node -> edge * children -> t -> t
  val fold_nodes :
    (node -> 'a -> 'a) -> t -> 'a -> 'a
  val fold :
    (node -> edge * children -> 'a -> 'a)
    -> t -> 'a -> 'a
  val harvest : t -> node -> t -> t
end
\end{verbatim}
Here \verb|type| declares an abstract data type and \verb|val|
declares values and functions, the latter just being values in a
functional programming language.  The type \verb|'a| is
polymorphic.  The actual signature in \textsc{O'Mega} contains additional
convenience functions that can be build from the functions presented
here.

Note that this implementation breaks the function~$\omega$ \eqref{eq:omega}
into products of functions~$\omega^0$ and~$\omega^1$, with
\begin{subequations}
  \begin{align}
    \omega_{n\mapsto\emptyset} &= \omega^0_{n} \\
    \omega_{n\mapsto\{(e_1,c_1),\ldots,(e_k,c_k)\}}
      &= \prod_{i=1}^k\omega^1_{n\mapsto(e_i,c_i)}\,.
  \end{align}
\end{subequations}
The function~$\omega^0$ (called \verb|add_node| here) can be used to
construct~$\bot_L\in\mathcal{D}_L$
from~$\epsilon\in\mathcal{D}_\emptyset$, while the action
of~$\omega^1$ (called \verb|add_offspring| here), does not leave
the category~$\mathcal{D}_L$.  This provides a better interface
for programming, but the~$\omega$ used in the main part of the paper
allowed a more concise writeup of the mathematical structures in
section~\ref{sec:DAG}. 

Correspondingly, the fold~$\Phi$ from~\eqref{eq:Phi} is broken into 
\verb|fold_nodes| processing all nodes and \verb|fold| processing
all~$N\to E\times C$ mappings element-by-element. The \texttt{ocaml}
equivalent of~\eqref{eq:copy-of-fold} is then
\begin{verbatim}
  let leaves' =
    fold_nodes add_node dag empty in
  fold add_offspring dag leaves'
\end{verbatim}
Note that this gives up some generality, because the~$\Phi$
from~\eqref{eq:Phi} could process the sets of~$E\times C$ as a whole
and not only element-by-element.  However, this interface is more
straightforward and is better tailored to our applications.

The function \verb|harvest| implements~$H$~\eqref{eq:harvest}. In particular,
\verb|harvest dag n dag'|
finds the subset of the DAG~\texttt{dag} that is reachable
from the node~\texttt{n} and adds it to the DAG~\texttt{dag'}.
This way, applications
can compute a minimal DAG for further processing.

Since the construction of the DAG~$\mathbf{D}_{\mathrm{P}}$
(cf.~section~\ref{sec:momenta}) is very simple, it had been combined
with the construction of~$\mathbf{D}_{\mathrm{F}}$
(cf.~section~\ref{sec:flavor}) in \textsc{O'Mega}~\cite{Moretti:2001zz}
before the structures described in this paper were elaborated.
However, the separation of the remaining stages described in
section~\ref{sec:feynman} forms the backbone of the current version of
\textsc{O'Mega}.

\subsection{Bundles}
\label{app:bundle-ocaml}

Here is the signature of the \texttt{Bundle}
module in \textsc{O'Mega}~\cite{Moretti:2001zz}.  Again a functor is applied to
the types~\verb|elt|, \verb|base| and the function~\verb|pi|,
corresponding to~$X$, $B$ and~$\pi$ respectively
\begin{verbatim}
module type Bundle = sig
  type elt
  type base
  val pi : elt -> base
  type t
  val empty : t
  val add : t -> elt -> t
  val inv_pi : t -> base -> fiber
  val base : t -> base list
end
\end{verbatim}
The semantics of the functions is evident from the discussion of
bundles in section~\ref{sec:bundles}.  Note that~$\pi$ is universal
for all bundles with this type, while~$\pi^{-1}$ depends on the
elements added to the bundle previously.

\end{appendices}

\bibliography{dags}
\end{document}